# Theory of Aces: Fame by chance or merit?

M.V. Simkin and V.P. Roychowdhury
*Department of Electrical Engineering, University of California, Los Angeles, CA 90095-1594*

**Abstract.** We study empirically how fame of WWI fighter-pilot aces, measured in numbers of web pages mentioning them, is related to their achievement or merit, measured in numbers of opponent aircraft destroyed. We find that on the average fame grows exponentially with achievement; to be precise, there is a strong correlation (~0.7) between achievement and the logarithm of fame. At the same time, the number of individuals achieving a particular level of merit decreases exponentially with the magnitude of the level, leading to a power-law distribution of fame. A stochastic model that can explain the exponential growth of fame with merit is also proposed.

An objective measure of achievement is difficult to define. As a result, the question of how fame is related to merit is ill posed. Fortunately, there is at least one case where an unquestionable measure of achievement does exist. This is the case of fighter-pilots, for whom achievement can be measured as a number of opponent aircraft destroyed. The website [1] contains names of all WWI fighter-pilot aces[1] together with the number of victories each of them had achieved.

These days there is an easily assessable index to fame: the number of web pages (as found using Google) that mention the person in question [2]. In the future we will refer to it as the number of *Google hits*. This approach is similar to measuring the impact of scientific papers in numbers of citations [3] and the importance of web pages in numbers of incoming hyperlinks [4]. Unfortunately, it is not enough to paste a name into Google search window and record the number of search results, because many aces have namesakes. For this reason search results were filtered to contain words like pilot or ace[2]. In many cases this method did not remove all of the namesakes and Google search results had to be visually inspected to complete the filtering. As it would be very time consuming to repeat this procedure for all 1,849 registered aces, the study was limited to include only German aces (392 total).

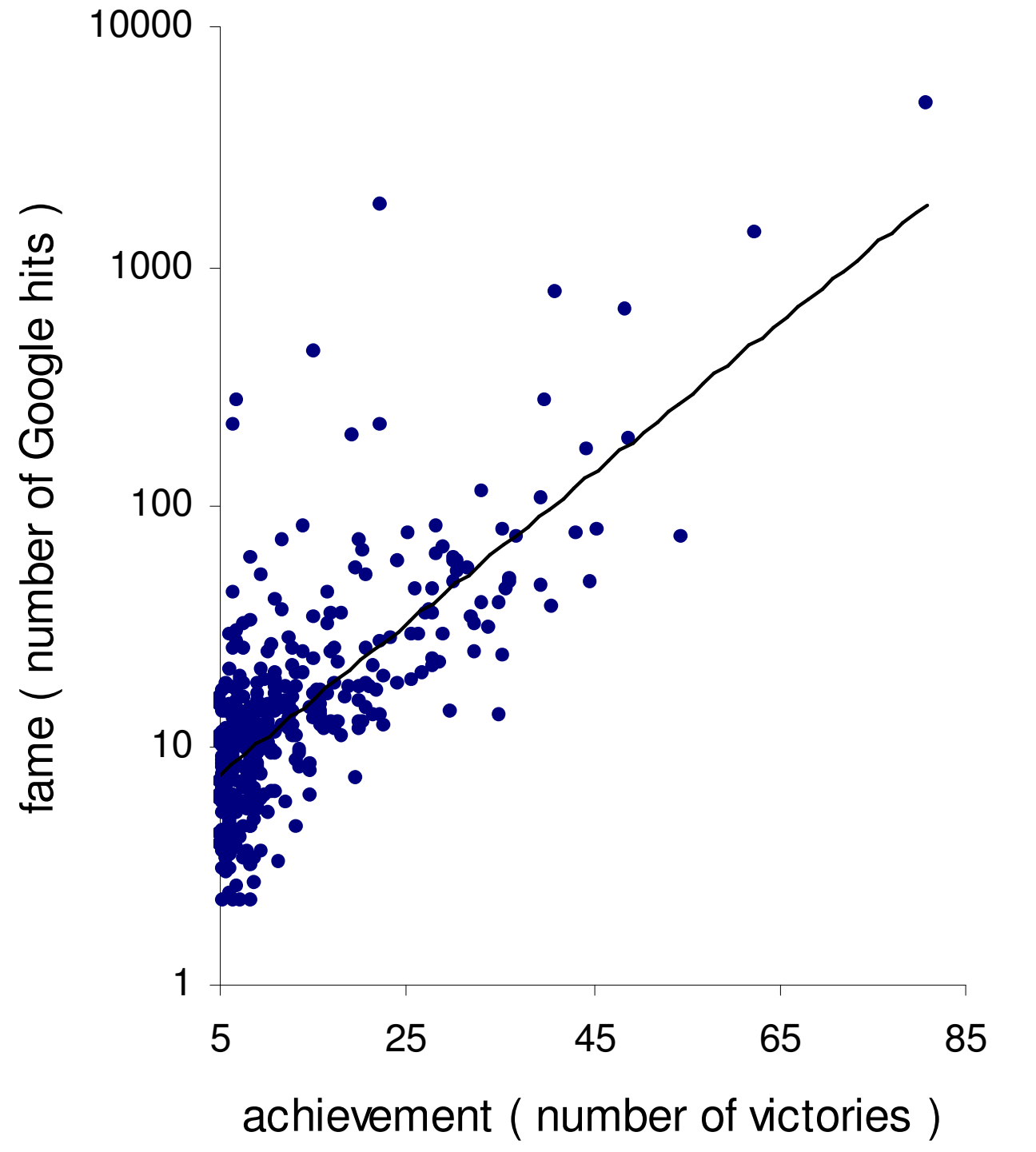

**Figure 1.** A scatter plot[3] of fame versus achievement for 392 German WWI aces. The correlation coefficient of 0.72 suggests that $0.72^2 \cong 52\%$ of the variation in fame is explained by the variation in achievement. The straight line is the fit using Eq.2 with $\beta \cong 0.074$.

[1] An ace is a fighter pilot who achieved five or more victories.
[2] The complete list of used filter words is: flying, pilot, ace, flieger, Jasta, Fokker, and WWI.

[3] There are many aces with identical values of both achievement and fame. Therefore for display purposes random numbers between zero and one were added to every value of achievement and fame. This way the scatter plot represents the true density of the data points.

Figure 1 is the result of this crusade. The correlation coefficient between achievement and the logarithm of fame is 0.72. In contrast the correlation between achievement and fame (without logarithm) is only 0.48. The significance of the correlation coefficient, $r$, is that $r^2$ is the fraction of variance in the data which is accounted for by linear regression. This means that about half ($0.72^2 \cong 52\%$) of difference in fame is determined by difference in achievement.

For a dozen of aces, achievement was not limited to the amount of won fights. For example, Max Immelmann is more known for his aerobatic maneuvers and Rudolf Stark for his books. Unfortunately, several aces acquired additional "fame" for less esteemed "achievements". This implies that the true correlation between achievement and fame is more than the above-mentioned value of 0.72.

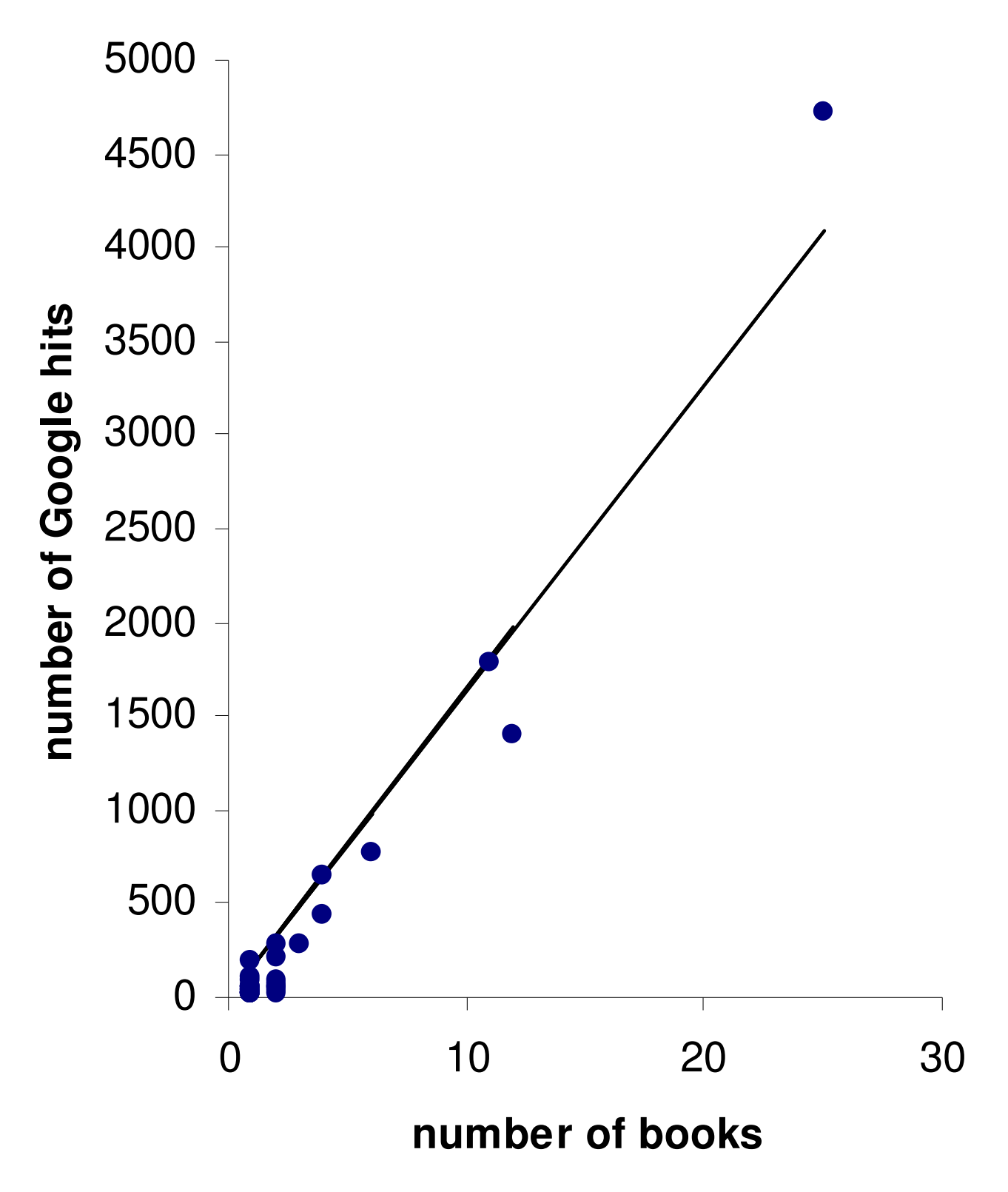

**Figure 2.** Number of Google hits versus number of books for 33 German WWI aces, who have at least one book written about them. The correlation coefficient of 0.985 suggests that $0.985^2 \cong 97\%$ of the variation in book-fame is explained by the variation in Google-fame. The data on the numbers of books were provided by F. Olynyk.

While fame is not perfectly correlated with achievement – different measures of fame do perfectly correlate between themselves. Some 33 German WWI aces were famous enough to have books written about them. Figure 2 shows the scatter plot of book-fame versus Google-fame. The correlation coefficient of 0.985 suggests that $0.985^2 \cong 97\%$ of the variation in number of Google-hits is explained by the variation in number of books.

The frequency distributions of achievement and fame are shown in Figures 3 and 4 correspondingly. One can see that achievement is distributed exponentially and that fame has a power-law tail.

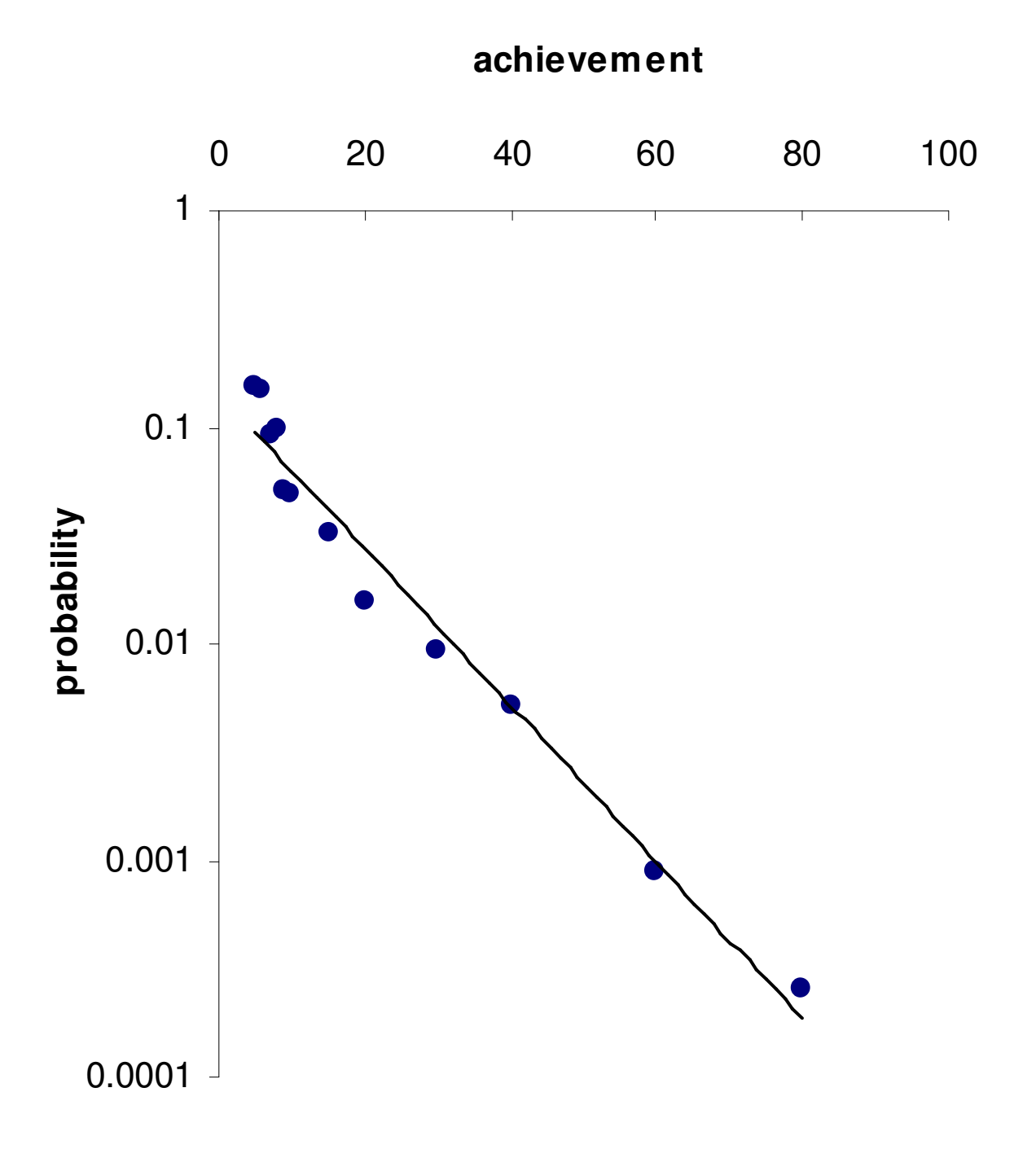

**Figure 3.** The distribution of achievement (number of victories) obtained using a sample of 392 German WWI aces. The straight line is the fit using Eq.1 with $\alpha \cong 0.083$.

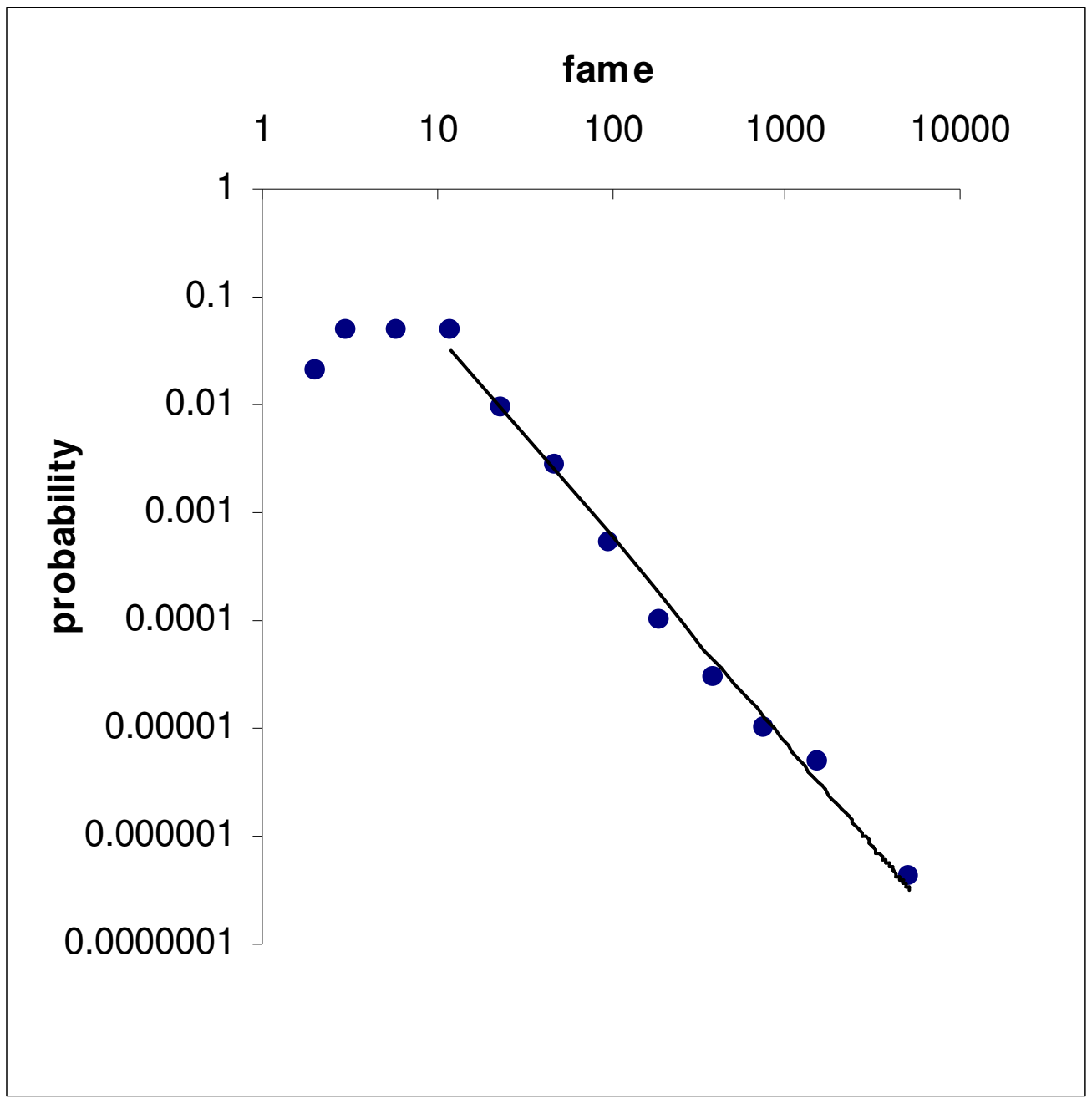

**Figure 4.** The distribution of fame (number of Google hits) computed using a sample of 392 German WWI aces. The straight line is the fit $p(F) \propto F^{-\gamma}$ with $\gamma \cong 1.9$.

When the distribution of achievement, $A$, is exponential,

$$p(A) = \alpha \exp(-\alpha \times A), \qquad (1)$$

and fame, $F$, grows exponentially with achievement,

$$F(A) = \exp(\beta \times A), \qquad (2)$$

then fame is distributed according to a power law. Clearly, elimination of $A$ from Eq. (1) using Eq. (2):

$$A(F) = \frac{1}{\beta}\ln(F);\; dA = \frac{dF}{\beta \times F}$$

leads to:

$$p(F) = \frac{\alpha}{\beta}F^{-\gamma}; \qquad \gamma = 1 + \frac{\alpha}{\beta}. \qquad (3)$$

Substituting the values $\alpha \cong 0.083$ and $\beta \cong 0.074$, obtained from the least-square fits of the data (see Figs. 1 and 3 into Eq.(3) we get $\gamma \cong 2.1$ which is quite close to $\gamma \cong 1.9$ obtained by fitting the actual distribution of fame (see Fig.4.

Exponential growth of fame with achievement leads to its unfair distribution. With 80 confirmed victories Manfred von Richthofen is the top-scoring ace of the WWI. With 4,720 Google hits[4] he is also the most famous. The total amount of opponent aircraft destroyed by German aces in WWI is 5050. At the same time there are 17,674 Google hits for all of the German aces. This means that Manfred von Richthofen accumulated $\frac{4{,}720}{17{,}674} \cong 27\%$ of fame, while being personally responsible for shooting down only $\frac{80}{5050} \cong 1.6\%$ of opponent aircraft. On the opposite side 60 lowest scoring aces (with 5 victories each) together shot down 300 aircraft, or $\frac{300}{5050} \cong 5.9\%$ of all aircraft destroyed. However, together they got only 463 Google hits, or $\frac{463}{17{,}674} \cong 2.6\%$ of fame.

A simple stochastic model can explain why fame grows exponentially with achievement. It is convenient to describe the dynamics of fame in terms of memes [5] (we use this word in the sense of a piece of information, which can pass from one mind to another). We define the fame of X as the number of people who know X, or, in other words, the number of memes about X. In practice we can't count the number of memes, but we can count the number of webpages. It is natural to assume that the number of webpages, mentioning X, is proportional to the number of memes.

The rate of encountering memes about X is obviously proportional to the current number of such memes in the meme pool. We will assume that when someone meets a meme about X, the probability that it will replicate into his mind is proportional to X's achievement (which thus plays the role of meme's Darwinian fitness). The rate of the spread of a meme about someone with achievement $A$ is thus:

$$s = \nu A\;. \qquad (5)$$

Here $\nu$ is an unknown independent of $A$ coefficient, which comprises the effects of all factors other than achievement on meme spread. The expectation value of the number of memes obeys the following evolution equation:

$$\frac{d\langle F\rangle}{dt} = s\langle F\rangle = \nu A\langle F\rangle\;. \qquad (6)$$

If at time 0 there was only one copy of the meme the solution of Eq. (5) is

$$\langle F(t)\rangle = \exp(\nu t \times A), \qquad (7)$$

[4] The data used in the paper were collected around May 2003. Today's numbers of Google hits are different.

which is Eq.(2) with

$$\beta = \nu t\ . \qquad (8)$$

In case of the aces $t$ is time passed since WWI. After substituting Eq.(8) into Eq.(3), we get:

$$\gamma = 1 + \frac{\alpha}{\nu t}$$

The value of $\gamma$ is consistent with the experimental data on aces if $\nu t \approx \alpha$.

Eq. (7) gives the expectation value of fame for a particular value of achievement. One can also ask what the distribution of fame is when the value of achievement is fixed. The probability that after time $t$ there is just one copy of the meme is:

$$p_1(t) = \exp(-st).$$

To have exactly 2 copies of the meme at time $t$, the meme should have replicated at some intermediate time, $\tau$, and afterwards two resulting memes should not replicate during time $t$-$\tau$. The probability of this is:

$$p_2(t) = \int_0^t sp_1(\tau)\exp(-2s(t-\tau))d\tau =$$

$$\exp(-st)(1-\exp(-st))$$

It can be verified by induction that the probability to have $F$ copies of the meme at time $t$ is:

$$p_F(t) = \exp(-st)(1-\exp(-st))^{F-1} \qquad (9)$$

After substituting Equations (5) and (8) into Equation (9), we get:

$$p_F(A) = \exp(-\beta A)(1-\exp(-\beta A))^{F-1}\ . \qquad (10)$$

Using Eq.(10) we can obtain the exact formula for the distribution of fame in our model: see Appendix A.

The above derivation is almost identical to the one, used by Yule to explain the power law in the frequency distribution of sizes of biological genera. Yule [6] considered two types of mutations: specific, which lead to a new specie and occur at rate $s$, and generic, which leads to a new genus and occur at rate $g$. The total number of genera grows with time as $\exp(g \times t)$. Therefore the number of genera of age $t$ is proportional to $\exp(-g \times t)$. The number of species in a genus of age $t$ is $\propto \exp(s \times t)$. The problem is identical to ours with achievement, $A$, replaced with age of a genus, $t$, and factors $\alpha$ and $\beta$ with $g$ and $s$ correspondingly. Therefore it follows from Eq. (3) that the exponent of the Yule's distribution is $\gamma = 1 + g/s$. Apart from Yule's process our findings are relevant to some problems studied recently in the context of evolving networks [7] (see Appendix B).

According to Eq.(10), the probability to get a fame, which is ten times more than what you can expect for your achievement is as small as $e^{-10} \approx 5 \times 10^{-5}$, while the probability to get it ten times less is as big as 1/10. This should give a confidence to the famous and relief to the unknown.

## Appendix A

To obtain the distribution of fame we integrate the product of the distribution of achievement (Eq.(2)) and conditional probability of fame given the value of achievement (Eq.(8)):

$$p(F) = \int_0^\infty dA p_F(A) p(A) =$$

$$\int_0^\infty dA \exp(-\beta A)(1-\exp(-\beta A))^{F-1} \alpha \exp(-\alpha A) =$$

$$\frac{\alpha}{\beta}\int_0^1 dx x^{\frac{\alpha}{\beta}} (1-x)^{F-1} = \frac{a}{\beta}\mathrm{B}\left(\frac{\alpha}{\beta}+1, F\right) =$$

$$\frac{\alpha}{\beta}\frac{\Gamma\left(\frac{\alpha}{\beta}+1\right)\Gamma(F)}{\Gamma\left(\frac{\alpha}{\beta}+1+F\right)}$$

Here $\mathrm{B}$ and $\Gamma$ are Euler's Beta and Gamma functions. Using known properties of $\Gamma$ function one can obtain the large $F$ asymptotic of the above:

$$p(F) \propto \frac{\alpha}{\beta}\frac{\Gamma\left(\frac{\alpha}{\beta}+1\right)}{F^{\frac{\alpha}{\beta}+1}}\ .$$

## Appendix B

The problem of finding distribution of fame when the value of achievement is fixed is identical to that of finding degree distribution of nodes of the *same age* in a growing network [8]. In section IV.A of Ref. [8] this problem was solved and the distribution was found to be exponential in agreement with our results. Note that our derivation (which followed that of Yule [6]) is simpler.

The problem of finding distribution of fame for all values of achievement is identical to that of finding degree distribution of nodes of the *same age* in a growing network, in which the nodes have exponentially distributed multiplicative fitness. Such network was studied using numerical simulations in Ref. [9], but no analytical results were obtained. One can use results of this paper to show that the degree distribution of nodes of the *same age* in such network obeys a power law.

Our results are also relevant to theory of citing. Redner [10] observed that the distribution of citations to papers *published during one year* and cited next seventeen years follows a power law. This observation can not be explained by a cumulative advantage model [6], [11], [12], in which the rate of citing a particular paper is proportional to the number of citations it had already received. Krapivsky and Redner [8] have shown that cumulative advantage leads to exponential distribution of citations to papers of the same age. The results of the present paper suggest that Redner's observation can be explained by a cumulative advantage model with multiplicative fitness [9] if we assume that the distribution of fitness is exponential. One may speculate that this fitness is related to merit, as it was in the case of aces.